\documentclass{article}
\usepackage[numbers,sort&compress,square]{natbib}
\usepackage{dcase2021,graphicx,url,times}
\usepackage{amsmath, amsthm, amssymb, amsfonts}
\usepackage{comment}
\usepackage{booktabs}
\usepackage{siunitx}
\usepackage{tabularx}
\usepackage{bm}
\usepackage{placeins}
\usepackage{algorithm}
\usepackage{algpseudocode}
\usepackage{enumitem}
\usepackage{siunitx}

\usepackage{multicol, multirow}
\usepackage{booktabs}
\usepackage[nameinlink]{cleveref}
\usepackage{caption}
\usepackage{subcaption}
\usepackage{jabbrv}

\usepackage{lipsum}

\usepackage{etoolbox}           
\renewcommand{\bfseries}{\fontseries{b}\selectfont} 
\robustify\bfseries             
\newrobustcmd{\B}{\bfseries}    

\newcommand{\etal}{ et al.\ }
\newcommand{\V}{$\times$}
\newcommand{\X}{$\checkmark$}

\newcommand{\LE}{$\text{LE}_\text{CD}$}
\newcommand{\LR}{$\text{LR}_\text{CD}$ }

\newcolumntype{C}{>{\centering\arraybackslash}X}

\title{What Makes Sound Event Localization and Detection Difficult?\\Insights from Error Analysis}

\name{
    Thi Ngoc Tho Nguyen$^{1}
        \thanks{This research was supported by the Singapore Ministry of Education Academic Research Fund Tier-2, under research grant MOE2017-T2-2-060.}$,
	Karn N. Watcharasupat$^{1}
	    \thanks{K. N. Watcharasupat acknowledges the support from the CN Yang Scholars Programme, Nanyang Technological University, Singapore.}$,
} \secondlinename{	  
	Zhen Jian Lee,
	Ngoc Khanh Nguyen,
    Douglas L. Jones$^{2}$, 
    Woon Seng Gan$^{1}$
}

\address{
    $^1$ School of Electrical and Electronic Engineering, Nanyang Technological University, Singapore.\\          
    $^2$ Dept. of Electrical and Computer Engineering, University of Illinois at Urbana-Champaign, IL, USA. \\
    \{nguyenth003, karn001\}@e.ntu.edu.sg, zhenjianlee@gmail.com, \\ ngockhanh5794@gmail.com, dl-jones@illinois.edu, ewsgan@ntu.edu.sg
}

\begin{document}

\ninept
\maketitle

\begin{sloppy}

\begin{abstract} 
Sound event localization and detection (SELD) is an emerging research topic that aims to unify the tasks of sound event detection and direction-of-arrival estimation. As a result, SELD inherits the challenges of both tasks, such as noise, reverberation, interference, polyphony, and non-stationarity of sound sources. Furthermore, SELD often faces an additional challenge of assigning correct correspondences between the detected sound classes and directions of arrival to multiple overlapping sound events. Previous studies have shown that unknown interferences in reverberant environments often cause major degradation in the performance of SELD systems. To further understand the challenges of the SELD task, we performed a detailed error analysis on two of our SELD systems, which both ranked second in the team category of DCASE SELD Challenge, one in 2020 and one in 2021. Experimental results indicate polyphony as the main challenge in SELD, due to the difficulty in detecting all sound events of interest. In addition, the SELD systems tend to make fewer errors for the polyphonic scenario that is dominant in the training set. 
\end{abstract}



\section{Introduction}
\label{sec:intro}

Sound event localization and detection (SELD) has many applications in urban sound sensing~\cite{Salamon2017Cnn}, wildlife monitoring~\cite{Stowell2016bird}, surveillance~\cite{Foggia2016Surveillance}, autonomous driving~\cite{nandwana2016car}, and robotics~\cite{valin2004localization}. SELD is an emerging research field that aims to combine the tasks of sound event detection (SED) and direction-of-arrival estimation (DOAE) by jointly recognizing the sound classes, and estimating the directions of arrival (DOA), the onsets, and offsets of detected sound events~\cite{Adavanne2019seld}.

The introduction of the SELD task in the 2019 Challenge on Detection and Classification of Acoustic Scenes and Events (DCASE) has significantly accelerated SELD research. Many significant contributions have been made over the last three years in terms of datasets, evaluation metrics, and algorithms~\cite{politis2020overview}. The TAU Spatial Sound Events dataset~\cite{adavanne2019dcasedataset} used in DCASE 2019 included only stationary sound sources, with 72 room impulse responses (RIRs) from 5 different locations, and only 20 distinct samples for each of the 11 sound classes. The TAU-NIGENS Spatial Sound Events dataset~\cite{politis2020dcasedataset} used in DCASE 2020 saw an introduction of moving sound sources, more RIRs from 15 different locations, and 14 sound classes extracted from the NIGENS General Sound Events Database~\cite{Trowitzsch2019TheDatabase}, with around 30 to 50 distinct samples per class. The 2021 edition~\cite{politis2021dcasedataset} introduced unknown directional interferences, making the sound scenes more realistic, in addition to the increase in the maximum polyphony of target events to three, from two in the 2019 and 2020 runs. The number of sound classes was reduced to 12, as some classes were used as interferences.
All three SELD datasets provide both first-order ambisonic (FOA) and microphone array (MIC) formats. 

The SELD evaluation metrics have evolved over the past three years. In DCASE 2019, SED and DOAE performances were evaluated independently. Segment-wise error rate (ER) and F1 score evaluation were used for SED~\cite{Mesaros2016_MDPI}, while frame-wise DOA error and frame recall were used for DOAE~\cite{adavanne2018doaRcnn}. Since 2020, SED and DOAE were evaluated jointly with location-dependent ER and F1 score for SED, and class-dependent localization error (LE) and localization recall (LR) for DOAE~\cite{mesaros2019seldeval}. The 2021 metrics further take into account overlapping same-class events~\cite{politis2021dcasedataset}.

On the algorithm aspect, there have been many developments for SELD, inside and outside the DCASE Challenges, in the areas of data augmentation, feature engineering, model architectures, and output formats. In 2015, an early monophonic SELD work by Hirvonen \cite{hirvonen2015classification} formulated SELD as a classification task, where each output class represents a sound class-location pair. In 2018, Adavanne\etal pioneered a seminal polyphonic SELD work that used a single-input multiple-output convolutional recurrent neural network (CRNN) model, SELDnet, to jointly detect sound events and estimate the corresponding DOAs~\cite{Adavanne2019seld}. In 2019, Cao\etal proposed a two-stage strategy by training separate SED and DOA models~\cite{cao2019polyphonic}, then using the SED outputs as masks to select the DOA outputs, significantly outperforming the jointly-trained SELDnet. Cao\etal later proposed an end-to-end SELD network~\cite{cao2020ein} that used soft parameter sharing between the SED and DOAE encoder branches and output trackwise predictions. An improved version of this network~\cite{cao2021ienv2} replaced the bidirectional gated recurrent units (GRU) with multi-head self-attention (MHSA) to decode the SELD outputs~\cite{cao2021ienv2}. In 2020, Shimada\etal proposed a new output format for SELD which unified SED and DOAE into one loss function~\cite{shimada2021accdoa}. This was amongst the few works which successfully used the linear-frequency for spectrograms and interchannel phase differences as input features, instead of the mel spectrograms. A new CNN architecture, D3Net \cite{Takahashi2021}, was adapted into a CRNN for this work and showed promising results. In another research direction, Nguyen\etal proposed to solve SED and DOAE separately, use a bidirectional GRU to match the SED and DOAE output sequences, then produce event-wise SELD outputs~\cite{tho2020smn, Nguyen2021ANetwork}. This was based on the observation that different sound events often have different onsets and offsets, resulting in temporal matching in the SED and DOAE output sequences. In 2021, Nguyen\etal proposed a new input feature, SALSA, which spectrotemporally aligns the spatial cues with the signal power in the linear-frequency scale to improve SELD performance~\cite{Nguyen2021DCASEDetection}. 

The top SELD system for DCASE 2019 trained four separate models for sound activity detection, SED, single-source DOAE, and two-source DOAE~\cite{Kapka2019seld}. The top systems for both DCASE 2020 and 2021 synthesized a larger dataset from the original data, employed many data augmentation techniques, and combined different SELD models into ensembles~\cite{du2020dcasetop, Shimada2021EnsembleDetection}. Other highly ranked solutions also intensively used data augmentation and ensemble methods.

Since SELD consists of both SED and DOAE tasks, it inherits many challenges from both SED and DOAE, such as noise, reverberation, interference, polyphony, and non-stationarity of sound sources. Furthermore, SELD often faces an additional challenge in correctly associating SED and DOAE outputs of multiple overlapping sound events. In an attempt to dissect the difficulties of the SELD task, Politis\etal compared the performances of the same SELD system in different acoustic environments~\cite{politis2021dcasedataset} with different combinations of noise, reverberation, and unknown interferences. The authors founded that, in absence of unknown interferences, ambiance noise has little negative effects on SELD performance, while reverberation significantly reduces the SELD performance in all noise combinations. Unknown interferences degrade SELD performances by the largest margin compared to noise and reverberation. In addition, using the FOA format generally produces better performance than the MIC format. 

To further understand the challenges facing SELD, we performed detailed error analysis on the SELD outputs, with the focus on polyphony, moving source, class-location interdependence, class-wise performance, and DOA errors, using our two SELD systems which both ranked second in the team category for the 2020 and 2021 DCASE Challenges~\cite{Nguyen2020EnsembleTracking, Nguyen2021DCASEDetection}. Experimental results showed that polyphony is the main factor that decreases the SELD performance across all the evaluation metrics, explaining why unknown interferences reduced the SELD performance by the largest extent. Interestingly, we also found that SELD systems do not necessarily favor single-source scenarios, which is easier than polyphonic cases. Instead, SELD systems achieved lower error rates in polyphonic cases which dominate the training dataset. The rest of the paper is organized as follows. \Cref{sec:method} describes our analysis method. \Cref{sec:results} presents the experimental results and discussions. Finally, we conclude the paper in \Cref{sec:concl}.

\section{Analysis Method}
\label{sec:method}
In this section, brief descriptions of the SELD datasets and systems are provided. Error analyses were performed on the SELD outputs of the two SELD systems which both ranked second in the team category for the 2020 and 2021 DCASE Challenges~\cite{Nguyen2020EnsembleTracking, Nguyen2021DCASEDetection}. The 2021 version of the evaluation metrics was used in all analyses. For convenience, the TAU-NIGENS Spatial Sound Events 2020 and 2021 datasets used in the DCASE Challenges \cite{politis2020dcasedataset, politis2021dcasedataset} are referred to here as the SELD 2020 and 2021 datasets, respectively.

\subsection{Dataset}

\begin{table}[t]
    \centering
    \noindent\begin{tabularx}{\columnwidth}{lCC}
    \toprule
    Characteristics & 2020 & 2021 \\ 
    \midrule
    Channel format                  & FOA   & FOA \\
    Moving sources                  & \X    & \X  \\
    Ambiance noise                  & \X    & \X  \\
    Reverberation                   & \X    & \X  \\
    Unknown interferences           & \V    & \X  \\
    Maximum degree of polyphony     & 2     & 3   \\ 
    Number of target sound classes  & 14    & 12  \\
    Evaluation split                & eval  & test \\
    \bottomrule
    \end{tabularx}
    \caption {Comparison between 2020 and 2021 SELD datasets}  
    \label{tab: datasets}
\end{table}

Table~\ref{tab: datasets} summarizes some differences between the two SELD datasets. Since both of the SELD systems require the FOA format, only the FOA subset of the datasets were used in our experiments. Each of the dataset consists of \num{400}, \num{100}, \num{100}, and \num{200} one-minute audio recordings for the train, validation, test, and evaluation splits respectively. The azimuth and elevation ranges are $[\SI{-180}{\degree}, \SI{180}{\degree})$ and $[\SI{-45}{\degree}, \SI{45}{\degree}]$, respectively. 
During the developmental stage, the validation set was used for model selection while the test set was used for evaluation. During the evaluation stage, the train, validation, and test data (collectively known as the development split) were used for training evaluation models. For the 2020 SELD dataset, the results on the evaluation split were used for the error analyses. Since the ground truth for the evaluation split of the 2021 SELD dataset has not been publicly released at the time of writing, the results on the test split of the 2021 SELD dataset were used for error analysis instead.

\subsection{Evaluation metrics}

To evaluate the SELD performance, we used the official SELD evaluation metrics~\cite{politis2020overview} from the DCASE 2021 Challenge. The metrics not only jointly evaluate SED and DOAE, but also take into account the cases where multiple instances of the same class overlap. The SELD evaluation metrics consist of location-dependent error rate ($\text{ER}_{\le T}$) and F1 score ($\text{F}_{\le T}$) for SED; and class-dependent localization error ($\text{LE}_\text{CD}$), and localization recall ($\text{LR}_\text{CD}$) for DOAE. A sound event is considered a correct detection only if it has a correct class prediction and its estimated DOA is also less than $T$ away from the DOA ground truth, where $T=\SI{20}{\degree}$ for the official challenge. The DOAE metrics are also class-dependent, that is, the detected DOA is only counted if its corresponding detected sound class is correct. A good SELD system should have low $\text{ER}_{\le T}$, high $\text{F}_{\le T}$, low $\text{LE}_\text{CD}$, and high $\text{LR}_\text{CD}$. 

\subsection{SELD systems}

We denote two of our SELD systems that ranked second in the team categories of the 2020 and 2021 DCASE challenges as NTU'20 and NTU'21, respectively. Table~\ref{tab:seld_sys} shows the performances of the baselines, the top-ranked solutions, and our second-ranked systems in 2020 and 2021. NTU'20 is an ensemble of sequence matching networks~\cite{tho2020smn, Nguyen2020EnsembleTracking} while NTU'21 is an ensemble of different models trained on our new proposed SALSA features for SELD~\cite{Nguyen2021DCASEDetection}. Both systems use the class-wise output format, which can only detect a maximum of one event of a particular class at a time. Both systems outperformed the respective baselines by a large margin, and only perform slightly worse than the respective top-ranked system. The 2020 results in Table~\ref{tab:seld_sys} were computed using the 2020 SELD evaluation metrics. For subsequent sections, the results of the NTU'20 system were recomputed using the 2021 metrics. 

\begin{table}[t]
    \centering
    \footnotesize
    \noindent\begin{tabularx}{\columnwidth}{lX 
        S[
            detect-weight, 
            mode=text, 
            table-format=1.2]
            S[
            detect-weight, 
            mode=text, 
            table-format=1.3]
            S[
            detect-weight, 
            mode=text, 
            table-format=2.1]
            S[
            detect-weight, 
            mode=text, 
            table-format=1.3]}
    \toprule 
        Year & System &  
        $\text{ER}_{\le \SI{20}{\degree}}$ &
        $\text{F}_{\le \SI{20}{\degree}}$ &
        $\text{LE}_\text{CD}$ &
        $\text{LR}_\text{CD}$ \\
    \midrule
        2020 & 
            Baseline~\cite{politis2020dcasedataset} 
                    & 0.69 & 0.413 & 23.1\si{\degree} & 0.624 \\
        (eval)
            & \#1: USTC'20~\cite{du2020dcasetop}    
                    & \B0.20 & \B0.849 & \B6.0\si{\degree} & 0.885 \\
            & \#2: NTU'20~\cite{Nguyen2020EnsembleTracking}
                    & 0.23 & 0.820 & 9.3\si{\degree} & \B0.900\\ 
        \midrule
        2021 & 
            Baseline~\cite{politis2021dcasedataset} 
                    & 0.73 & 0.307 & 24.5\si{\degree} & 0.448 \\
        (test)
            & \#1: Sony'21~\cite{Shimada2021EnsembleDetection}
                    & 0.43 & 0.699 & \B11.1\si{\degree} & 0.732  \\
            & \#2: NTU'21~\cite{Nguyen2021DCASEDetection}     
                    & \B0.37 & \B0.737 & 11.2\si{\degree} & \B0.741 \\
    \bottomrule
    \end{tabularx}
    \caption{Performance of selected SELD systems.}
    \label{tab:seld_sys}
\end{table}

\section{Experimental results and discussion}
\label{sec:results}

In each subsection concerning a factor of variation, we performed an analysis on the data distribution of 2020 and 2021 SELD datasets, followed by an analysis of the SELD results. Overall, the 2021 dataset is much more challenging than the 2020 dataset. For detailed analyses, $\text{ER}_{\le T}$ is further broken down into substitution, deletion, and insertion errors, while $\text{F}_{\le T}$ is further broken down into precision and recall. Since the SELD metrics are segment-based, i.e., outputs are divided into segments of \SI{1}{\second} before being evaluated, we used the provided ground truth to group the segments based on polyphony (0, 1, 2, and 3 sources), static and moving sources to compute the metrics for each case. 

\subsection{Effect of polyphony}

\begin{figure}[t]
    \begin{minipage}[b]{.48\linewidth}
    \centering
    \centerline{\includegraphics[width=4.0cm]{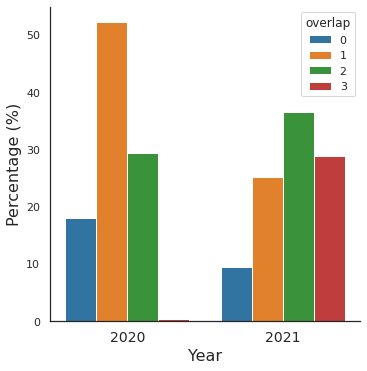}}
    \centerline{(a) Polyphonic distribution}\medskip
    \end{minipage}
    \hfill
    \begin{minipage}[b]{0.48\linewidth}
    \centering
    \centerline{\includegraphics[width=4.0cm]{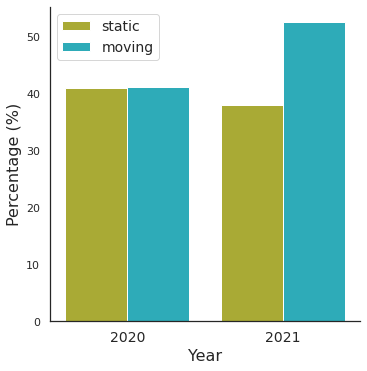}}
    \centerline{(b) Static vs moving}\medskip
    \end{minipage}
    \vspace{-1em}
    \caption{Segment-wise polyphonic and static distribution per year.}
    \label{fig:poly}
\end{figure}

\begin{table}[t]
    \setlength\tabcolsep{2pt}
    \footnotesize
    \centering
    \begin{tabularx}{\columnwidth}{
        X 
        *{7}{S[
            detect-weight, 
            mode=text, 
            table-format=2.3]
        }
    }
    \toprule
        &
        \multicolumn{3}{c}{2020} &
        \multicolumn{4}{c}{2021} \\
    \cmidrule(lr){2-4} \cmidrule(lr){5-8}
        Metrics & 
            \multicolumn{1}{c}{1} & 
            \multicolumn{1}{c}{2} & 
            \multicolumn{1}{c}{All} & 
            \multicolumn{1}{c}{1} & 
            \multicolumn{1}{c}{2} & 
            \multicolumn{1}{c}{3} & 
            \multicolumn{1}{c}{All} \\
    \midrule
        $\downarrow$ $\text{ER}_{\le \SI{20}{\degree}}$
            & \B0.108   
            & 0.331   
            & 0.232   
            & 0.349   
            & \B0.338   
            & 0.394   
            & 0.372 \\
        \quad$\downarrow$ Substitution   
            & \B0.029
            & 0.072   
            & 0.052   
            & \B0.093   
            & 0.104  
            & 0.129   
            & 0.114 \\
        \quad$\downarrow$ Deletion    
            & \B0.042   
            & 0.155   
            & 0.103   
            & \B0.091  
            & 0.137  
            & 0.182   
            & 0.152 \\
        \quad$\downarrow$ Insertion     
            & \B0.038   
            & 0.104   
            & 0.078   
            & 0.164   
            & 0.096   
            & \B0.083   
            & 0.105 \\
    \midrule
        $\uparrow$ $\text{F}_{\le \SI{20}{\degree}}$   
            & \B0.930   
            & 0.765   
            & 0.845   
            & \B0.784  
            & 0.763   
            & 0.704   
            & 0.737 \\
        \quad$\uparrow$ Precision  
            & \B0.932   
            & 0.788   
            & 0.875   
            & 0.757   
            & \B0.780   
            & 0.746  
            & 0.756 \\
       \quad$\uparrow$ Recall      
            & \B0.928   
            & 0.743   
            & 0.833   
            & \B0.813   
            & 0.747   
            & 0.666   
            & 0.719 \\
    \midrule
        $\downarrow$ $\text{LE}_\text{CD}$ 
            & \B5.6   
            & 13.4   
            & 9.4   
            & \B6.8   
            & 10.3   
            & 13.5   
            & 11.2 \\
        $\uparrow$ $\text{LR}_\text{CD}$ 
            & \B0.930   
            & 0.775   
            & 0.846   
            & \B0.816   
            & 0.764   
            & 0.701   
            & 0.741 \\
    \bottomrule
    \end{tabularx}
    \caption{SELD performance w.r.t. degree of polyphony}
    \label{tab:poly}
\end{table}

\Cref{fig:poly}(a) shows the segment-wise polyphonic distribution of 2020 and 2021 datasets, which are dominated by single-source and two-source segments, respectively. On average, there are \num{1.11} and \num{1.85} events per segment in the 2020 and 2021 datasets, respectively. \Cref{tab:poly} shows the breakdown of the SELD performance for each polyphonic case. The DOAE metrics clearly show that polyphony is a major cause of performance degradation. For both NTU'20 and NTU'21 systems, as the number of overlapping sources increases, \LE increases and \LR decreases. Interestingly, polyphony does not always degrade SED performance. The peak performances of $\text{ER}_{\le \SI{20}{\degree}}$ and precision were achieved in the degree of polyphony that dominates the respective dataset, which is single-source for the 2020 dataset and two-source for the 2021 dataset. This result suggests that one possible solution to tackle polyphony is to introduce more data samples for difficult cases. 

When the number of overlapping sources increases, the SED error compositions also change. The deletion error rate rapidly increases, the insertion error rate sharply decreases, and the substitution error rate increases. In addition, the recall rate decreases significantly. It is clear that the SELD systems struggle to detect all the present events in polyphonic cases. 

In the absence of any event of interest, the insertion error rates are \num{0.030} and \num{0.122} for NTU'20 and NTU'21 systems, respectively. When comparing the SELD performances between the 2020 and 2021 setups, the single-source results in 2021 are significantly worse than those in 2020 across all metrics. In addition, the substitution errors across all degrees of polyphony are much higher in the 2021 setup, than in 2020. These results show the detrimental effect of unknown interferences that were introduced in the 2021 dataset, consistent with the findings in~\cite{politis2021dcasedataset}. 

\subsection{Effect of moving sound sources}

\begin{table}[t]
    \setlength\tabcolsep{3pt}
    \footnotesize
    \centering
    \begin{tabularx}{\columnwidth}{X *{7}{S[
            detect-weight, 
            mode=text, 
            table-format=2.3]}}
    \toprule
        &
        \multicolumn{3}{c}{2020} &
        \multicolumn{3}{c}{2021} \\
    \cmidrule(lr){2-4} \cmidrule(lr){5-7}
        Metrics & 
        \multicolumn{1}{c}{Static} & 
        \multicolumn{1}{c}{Moving} & 
        \multicolumn{1}{c}{All} & 
        \multicolumn{1}{c}{Static} & 
        \multicolumn{1}{c}{Moving} & 
        \multicolumn{1}{c}{All} \\
    \midrule
        $\downarrow$ $\text{ER}_{\le \SI{20}{\degree}}$
            & \B0.214   & 0.239   & 0.232   
            & 0.379     & \B0.357 & 0.372 \\
        $\uparrow$ $\text{F}_{\le \SI{20}{\degree}}$   
            & \B0.854   & 0.841   & 0.845   
            & 0.731     & \B0.745   & 0.737 \\
        $\downarrow$ $\text{LE}_\text{CD}$ 
            & \B8.7   &  10.0   & 9.4   
            & \B10.5   & 11.7   & 11.2 \\
        $\uparrow$ $\text{LR}_\text{CD}$ 
            & \B0.847   & 0.846   & 0.846   
            & 0.725   & \B0.751   & 0.741 \\
    \midrule
        $\downarrow$ $\text{ER}_{\le \SI{180}{\degree}}$
            & \B0.166   & 0.168   & 0.171   
            & 0.334   & \B0.298   & 0.318 \\
        $\uparrow$ $\text{F}_{\le \SI{180}{\degree}}$   
            & \B0.898   & 0.891   & 0.892   
            & 0.778   & \B0.800   & 0.789 \\
    \bottomrule
    \end{tabularx}
    \caption{SELD performance of static and moving sources.}
    \label{tab:dynamic}
\end{table}

\Cref{fig:poly}(b) shows the segment-wise distribution of static and moving sound sources, not counting empty segments, based on the provided ground truth. A segment is considered a moving one if at least one sound source is moving. Since there are more overlapping sources in the 2021 dataset, the proportion of moving segments is significantly higher than the 2020 dataset. \Cref{tab:dynamic} presents the SELD performance for both cases. The $\text{LE}_\text{CD}$ of moving-source cases is higher than those of static-source cases, as expected. For the 2020 dataset, the $\text{LR}_\text{CD}$ are similar for both cases, and the performance gap for SED disappears when we compute location-independent SED metrics (by setting the DOA threshold to $T=180\si{\degree}$). These results suggest that moving sources have little effect on SED performance and mainly affect DOAE. For the 2021 dataset, all metrics are better for moving-source cases compared to single-source cases. This contradictory result may be due to the skewed distribution and requires further investigation once the evaluation ground truth is made available. 

\subsection{Class and location interdependency}

\begin{figure}[t]
    \centering
    \includegraphics[width=0.45\textwidth]{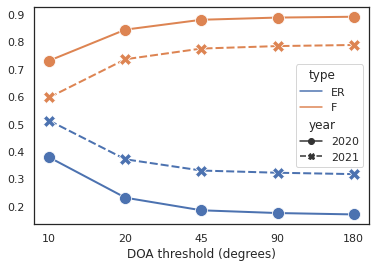}
    \vspace{-1em}
    \caption{SED performance across different DOA thresholds.}
    \label{fig:cloc}
\end{figure} 

\begin{figure}[t]
    \begin{minipage}[b]{.48\linewidth}
        \centering
        \centerline{\includegraphics[width=4.0cm]{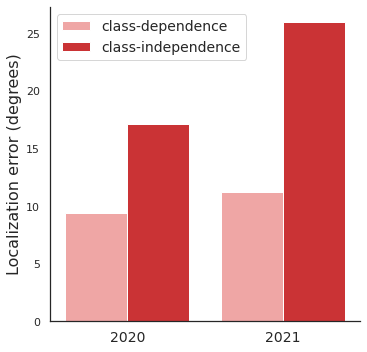}}
        \centerline{(a) Localization error}\medskip
    \end{minipage}
    \hfill
    \begin{minipage}[b]{0.48\linewidth}
        \centering
        \centerline{\includegraphics[width=4.0cm]{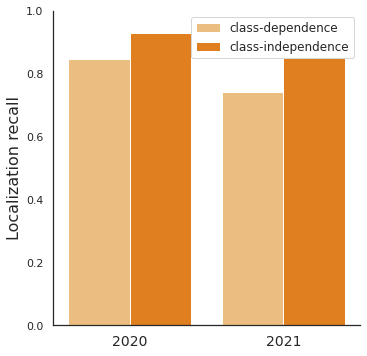}}
        \centerline{(b) Localization recall}\medskip
    \end{minipage}
    \vspace{-1em}
    \caption{Localization error and recall by class dependencies.}
    \label{fig:cloc2}
\end{figure}

To understand the dependency of location-dependent SED metrics on the correctness of the detected DOAs, we investigate the effect of the different DOA thresholds $T\si{\degree}$ on $\text{ER}_{\le T\si{\degree}}$ and $\text{F}_{\le T\si{\degree}}$, as shown in \Cref{fig:cloc}. The gaps between the SED metrics for $T=\SI{20}{\degree}$ and the location-independent $T=\SI{180}{\degree}$ are not significantly large, suggesting that many estimated DOAs are within the $\SI{20}{\degree}$ threshold. However, the location-dependent SED metrics deteriorate quickly as the DOA threshold reduces to $10\si{\degree}$, suggesting a significant number of the estimated DOAs deviate by more than $10\si{\degree}$ from the ground truth. 

To understand the dependency of classification-dependent DOA metrics on the correctness of the predicted classes, we show the classification-dependent and classification-independent LE and LD in \Cref{fig:cloc2}. When not accounting for the predicted class, the LR significantly increases, leading to some unwanted rise in LE.

\subsection{Class-wise performance}

\begin{figure}[t]
    \begin{minipage}[b]{.48\linewidth}
    \centering
    \centerline{\includegraphics[width=4.5cm]{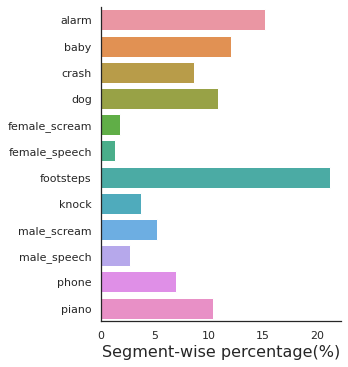}}
    \centerline{(a) Class distribution}\medskip
    \end{minipage}
    \hfill
    \begin{minipage}[b]{0.48\linewidth}
    \centering
    \centerline{\includegraphics[width=4.5cm]{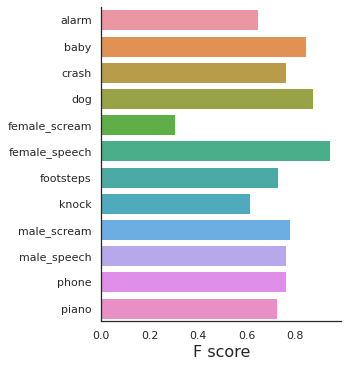}}
    \centerline{(b) Class-wise $\text{F}_{\le \SI{20}{\degree}}$}\medskip
    \end{minipage}
    \vspace{-1em}
    \caption{Segment-wise class distribution of 2021 SELD dataset (test split) and class-wise location-dependent F score of NTU'21 system.}
    \label{fig:class}
\end{figure}

Due to space constraints, we only included the segment-wise class distribution and the class-wise performance of 2021 setup in \Cref{fig:class}. The segment-wise class distribution in \Cref{fig:class}(a) is highly skewed, with the \emph{footstep} class accounting for the highest proportion of \SI{21.2}{\percent}, while the \emph{female speech} accounting for the lowest at \SI{1.3}{\percent}. However, the class-wise 
$\text{F}_{\le \SI{20}{\degree}}$ scores are more even, and the class with the highest segment-wise proportion does not correspond to highest $\text{F}_{\le \SI{20}{\degree}}$ score. One possible reason is that it is difficult to detect all \emph{footstep} sound due to discontinuities, low bandwidth, and low energy. In addition, class-wise performance is highly dependent on the SELD model and the quality of training samples. Interestingly, the \emph{female speech} class with the highest $\text{F}_{\le \SI{20}{\degree}}$ score of \SI{94.2}{\percent} has the lowest segment-wise proportion. Other classes such as \emph{knock} and \emph{male speech} also have high $\text{F}_{\le \SI{20}{\degree}}$ scores despite the low segment-wise proportions.

\subsection{Azimuth vs elevation error}

For the NTU'20 system, the $\text{LE}_\text{CD}$ contributed by azimuth and elevation are \SI{6.3}{\degree} and \SI{5.3}{\degree}, respectively. For the NTU'21 system, the $\text{LE}_\text{CD}$ contributed by azimuth and elevation are \SI{7.9}{\degree} and \SI{6.2}{\degree}, respectively. The azimuth and elevation errors are similar although the azimuth range of $[\SI{-180}{\degree}, \SI{180}{\degree})$ is much larger than elevation range of $[\SI{-45}{\degree}, \SI{45}{\degree}]$, suggesting that it is more difficult to estimate elevation angles than azimuth angles.

\section{Conclusion}
\label{sec:concl}
In realistic acoustic conditions with noise and reverberation, polyphony and unknown interferences appear to be the biggest challenges for SELD. In the presence of unknown interferences, SELD systems tend to make more substitution errors. When there are several sound events, either due to polyphony or unknown interferences, the SELD systems struggle to detect all events of interests, leading to low recall and high deletion error rate. Interestingly, the overall SED error rate is at the lowest for the polyphonic case that dominates the dataset. Moving sound sources mainly increase the localization errors, leading to small reduction in location-dependent SED metrics. High segment-wise representation of a class also does not necessary translate to high SED performances. Localization error reduction poses significant challenge beyond a threshold, especially as elevation errors are often as high as azimuth errors. The study of same-class polyphonic events is left for future works due to the limitations of the current systems studied.


\bibliographystyle{IEEEtran}
\bibliography{refs}

\end{sloppy}
\end{document}